\documentclass[11pt,twoside]{article}
\usepackage{asp2010}

\resetcounters

\begin{document}
\bibliographystyle{asp2010}

\title{Superluminal Waves and the Structure of Pulsar Wind Termination Shocks}

\author{Iwona Mochol and John G. Kirk
  \affil{Max-Planck-Institut f\"ur Kernphysik, Postfach 10 39 80, 69029 Heidelberg, Germany}
  }

\begin{abstract}
  The termination shock of a pulsar wind is located roughly where the 
  ram pressure matches that of the surrounding medium. 
Downstream of the shock, MHD  models of the diffuse nebular emission suggest the plasma is weakly   magnetized. However, the transition from a Poynting-dominated MHD
  wind to a particle-dominated flow is not well understood. We
  discuss a solution of this "$\sigma$-problem" in which a striped
  wind converts into a strong, superluminal electromagnetic wave. 
This mode slows   down as it propagates radially, and its ram pressure tends to a
  constant value at large radius, a property we use to match the
  solution to the surrounding nebula. The wave thus forms a pre-cursor to the 
termination shock, which occurs at the point where the wave dissipates. 
Possible damping and dissipation mechanisms are discussed qualitatively. 
\end{abstract}

\section{Introduction - MHD and the $\sigma$-problem}  

The electromagnetic emission from pulsars is only a small fraction of
their spin-down power, most of which must be carried away
by a relativistic wind, consisting of particles and low-frequency 
electromagnetic fields. Close to the light cylinder, the latter
component is expected to dominate the energetics, in the sense that the ratio
$\sigma$ of the energy flux carried by the fields to that carried by
the particles is much larger than unity. Numerical solutions of the 
force-free magnetosphere problem e.g., \cite{2012arXiv1205.0889P}
support the idea that the wind is launched in the form of magnetic stripes 
of alternating polarity frozen into the radial plasma flow 
\citep{1990ApJ...349..538C}. 

Locally, the physics of a radial wind depends on the energy flux density
it carries, which may be expressed
by the dimensionless parameter $a_{\rm L}=(e^2L/m^2c^5)^{1/2}$.
Formally,
$L$ is $4\pi$ times the luminosity per unit solid angle in a given
radial direction. In the simplest case this equals the spin-down power.  
Physically,
$a=a_{\rm L}r_{\rm L}/r$ is the strength parameter that 
a circularly polarized vacuum wave would need, in order to carry
the same energy flux density as the pulsar wind. ($r_{\rm L}=c/\omega$ is the 
radius of the light cylinder and $2\pi/\omega$ is the pulsar period. The 
strength parameter of a vacuum wave is defined as $a=eE/mc\omega$, with $E$ the 
electric field amplitude.)

But a pulsar is not surrounded by vacuum; the wave carries 
a finite particle flux. How many particles are available in
an outflow is determined by the pair production rate (multiplicity
coefficient $\kappa$) in the magnetosphere and can be quantified by the
ratio of the luminosity to the mass-loss rate (times $c^2$): 
$\mu=a_{\rm L}/4\kappa$
\citep{2001ApJ...547..437L, 2012ApJ...745..108A}, which equals the
Lorentz factor which each particle would have if the entire luminosity
were carried by the particles only.

When it encounters the surroundings, a stellar wind is decelerated and
terminates at an (approximately) standing shock, 
where its ram pressure is balanced
by the confining pressure of the medium. At the shock, the energy is
deposited in relativistic particles that are responsible for the
measured radiation. In the pulsar case, the problem with this scenario 
is that in an ideal, radial
ultrarelativistic MHD wind there is no plausible mechanism of
converting the Poynting flux into the particle energy flux.
This is often called the $\sigma$-problem.

A solution can be found only by looking beyond the ideal MHD description. One
possibility is a scenario in which the MHD wind converts into a strong
electromagnetic (EM) wave of superluminal phase speed before reaching the shock
\citep{1975Ap&SS..32..375U,1996MNRAS.279.1168M}. The new mode can be
thought of as a shock precursor \citep{2010PPCF...52l4029K}, since the 
point of conversion is causally connected to the external medium. These 
modes accelerate particles to relativistic energies in a plane transverse
to the direction of motion, and so they transfer most of the flow energy 
from the fields into the plasma. The mode conversion process itself, 
which is not considered here, can probably only be investigated using 
two-fluid or PIC simulations. However, just like an MHD shock, it 
is constrained by jump conditions that follow from the
parameters of the MHD wind and the pressure of the external medium.

\section{And Beyond MHD}

As EM waves can propagate only in an underdense plasma, mode
conversion can happen only beyond a certain distance from the star
$r>r_{\rm c}$, where, due to spherical expansion, the particle density 
drops below a critical value. When this happens, a fraction of
the flow energy is available to the transverse degrees of freedom. 
The critical radius, expressed in terms of the pulsar wind parameters 
is $r_{\rm c}\approx (a_{\rm L}/\mu)r_{\rm L}=4\kappa r_{\rm L}$
\citep{2012ApJ...745..108A}. 
When a wave is launched far outside this cut-off distance, it
resembles a large amplitude vacuum wave, but close to the cut-off the
plasma strongly affects the wave properties and a self-consistent solution
deviates from a vacuum wave.

The feature which distinguishes strong waves from linear EM waves
is that they are able to drive particles to extremely
relativistic energy $\gamma\approx a$ in only half a period. 
To describe the propagation of a
strong plane wave in a plasma one has to solve the full nonlinear set of
equations of particle motion coupled to Maxwell equations
\citep{akhiezerpolovin56, 1971PhRvL..27.1342M}. In this self-consistent
approach particles are not test particles; their conduction
currents contribute to maintaining the wave
fields. Since EM waves have a nonvanishing electric field even in the 
local fluid frame, they are excluded from an MHD description.
In pulsar winds, the simplest description that includes them
is a cold, two-fluid ($e^{\pm}$) plasma. 
A monochromatic solution of these equations 
can be found that describes a
circularly polarized EM wave, propagating in a plasma with superluminal phase
velocity, but subluminal group speed $c\beta_{*}$. 
In it, the electron and positron fluids move with equal
parallel momenta $p_{\parallel}$, but have equal amplitude, oppositely directed
oscillations in transverse momenta $p_{\perp}$, which is  
everywhere perpendicular
to the electric field. This generates a conduction current, that, in the 
frame in which the wave has zero group speed, exactly
balances the displacement current.
In the general case, the wave group speed does not coincide with the parallel 
component of the fluid 3-velocity, so that there is a nonvanishing 
particle flux in the wave frame. 

Under pulsar conditions the wave is expected to be radial. At distances
$r\gg r_{\rm L}$ it is, to a first approximation, plane, and the 
deformation due to spherical geometry can be treated using 
perturbation analysis, expanding the relevant equations in
the small parameter $\epsilon=r_{\rm L}/r\ll1$. The first-order equations
describe the radial evolution of the phase-averaged
quantities associated with the zeroth-order plane wave. 
These equations are the continuity equation, the energy
conservation equation, and an equation for 
the evolution of the radial momentum flux. In contrast to the MHD wind,
the radial momentum flux is not conserved in spherical geometry
for the EM modes. However,
it can be shown \citep{mocholphd} that the third integral of motion 
for both
circularly and linearly polarized modes is the 
phase-averaged Lorentz factor of the particles, measured in the
laboratory frame $\left\langle{\gamma_{\rm lab}}\right\rangle$.   

\begin{figure}[ht]
\begin{minipage}[b]{0.5\linewidth}
\centering
\includegraphics[angle=270,width=1.0\textwidth]{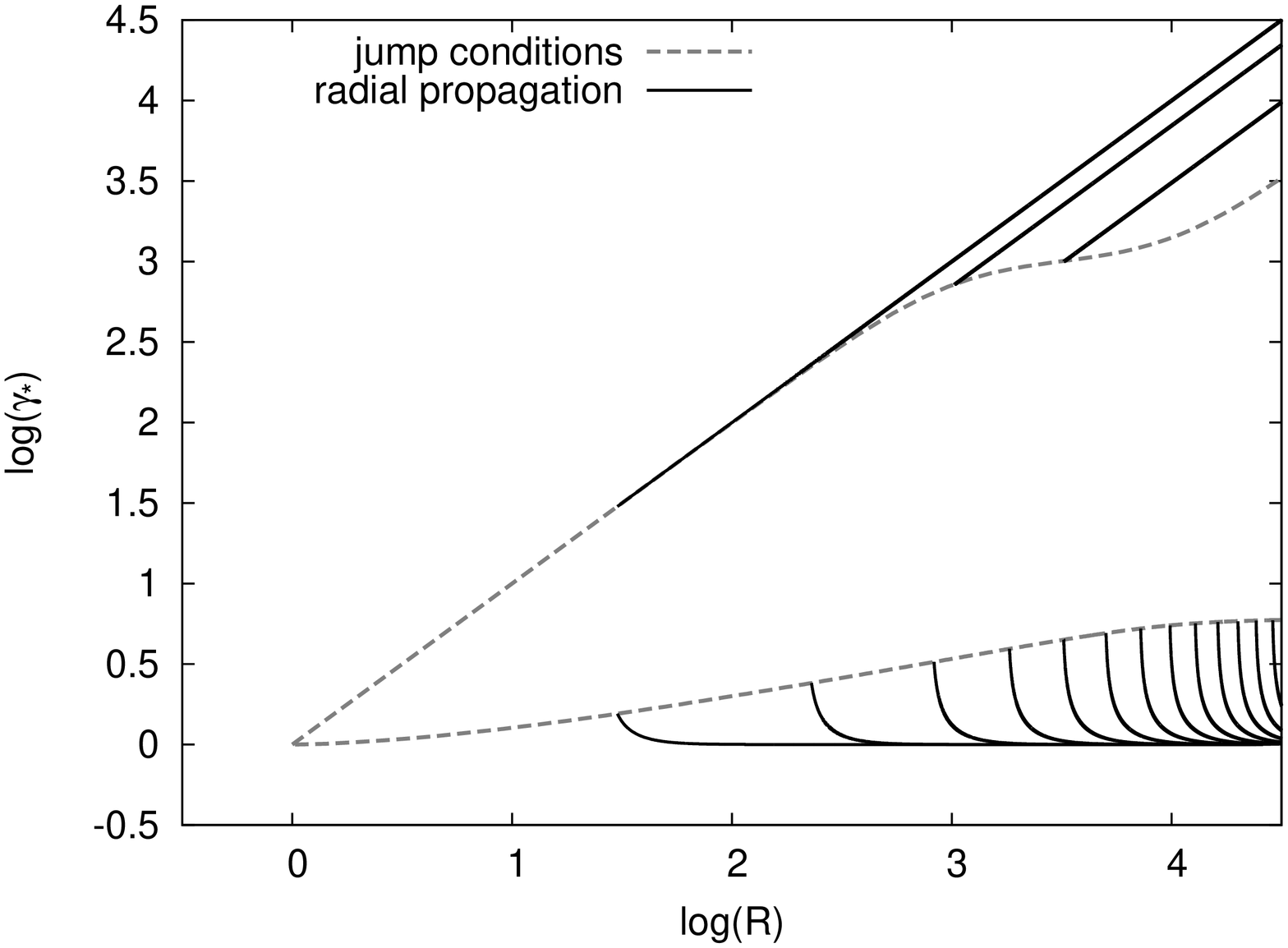}
\caption{The Lorentz factor of a circularly polarized strong wave, 
corresponding to an MHD wind with 
$\mu=10^4$, $\sigma=100$, $a_{\rm L}=3.4\times10^{10}$.}
\label{fig:figure1ag}
\end{minipage}
\begin{minipage}[b]{0.5\linewidth}
\centering
\includegraphics[angle=270,width=1.0\textwidth]{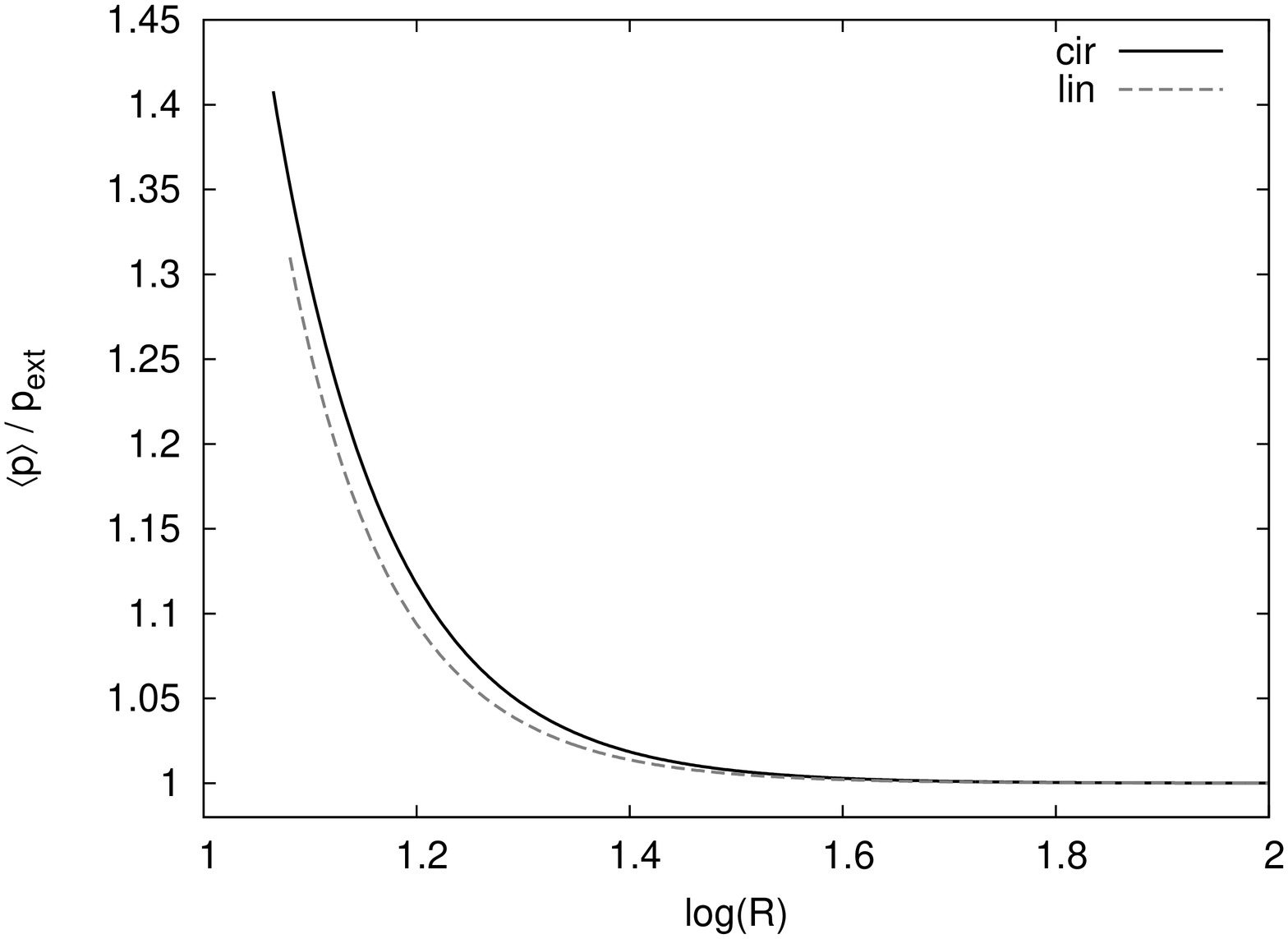}
\caption{Pressure of a confined mode (circularly and linearly polarized) as a function of rescaled radius $R=\left(r/r_{\rm L}\right)\left(\mu/a_{\rm L}\right)$.}
\label{fig:figure2ag}
\end{minipage}
\end{figure}

To find the initial condition, one has to solve jump conditions
between the MHD and the EM wave, to ensure that 
they carry the same particle,
energy and radial momentum fluxes \citep{2010PPCF...52l4029K,
  2012ApJ...745..108A}.  In Fig.~\ref{fig:figure1ag} we show the Lorentz
factor of an EM strong wave $\gamma_{*}=\left(1-\beta_{*}^2\right)^{-1/2}$, 
obtained from the jump conditions (dashed curves), and its radial 
evolution (solid curves) for
different launching points. There are two solutions of the jump
conditions that describe two possible EM modes: a free-escape mode
(higher branch) and a confined mode (lower branch). Their behaviour is
very different: at large distances the free-escape wave accelerates
whereas the confined one decelerates. Keeping in mind that the wind
solution should be matched to the slowly expanding
nebula, we concentrate only on the confined mode. The radial dependence of
its ram pressure is shown in Fig.~\ref{fig:figure2ag}, for both linear
and circular polarizations.

Since the ram pressure of the confined mode tends to a constant value at
large radius, we are able to find an unique solution that matches
asymptotically a given pressure $p_{\rm ext}$ of the external medium. In
fact, to constrain a wave at launch uniquely, four quantities have to
constrained. These are: the conversion
radius $R_0$, initial group speed $\beta_{*0}$ of a wave and initial
particle momenta $p_{\parallel 0}$, $p_{\perp 0}$. The jump conditions define
three of them, leading to the red curves in Fig.~\ref{fig:figure1ag}, and 
the external pressure can be
used to determine the fourth one --- $R_0$. 
The existence of the third integral of motion 
$\left\langle {\gamma_{\rm lab}} \right\rangle$ makes this task easy, and leads to a 
unique stationary solution for the shock precursor for given 
MHD wave parameters $\sigma$, $\mu$, $a_L$, and external pressure $p_{\rm ext}$.

\section{Damping and Shock Formation}

Asymptotic matching of the ram and external pressures leads to a 
self-consistent solution for the wave into which a given MHD wind 
converts. However, EM waves are also damped, 
because the accelerated particles they contain emit photons 
\citep{1971ApJ...165..523G,1978A&A....65..401A}
or scatter pre-existing photons from external sources.
%The former is called  nonlinear inverse Compton (NIC) scattering; 
%if the wave is strong, the emitted
%spectrum resembles that of synchrotron radiation. 
Damping 
removes two of the integrals of motion, leaving only particle flux conserved, 
and the resulting system must be integrated numerically.

However, it has been shown both analytically
\citep{1973PhFl...16.1480M,1978JPlPh..20..313L} and numerically
\citep{1978JPlPh..20..479R} that strong waves are unstable to small
density perturbations in the direction of motion, provided the particles
stream through the wave sufficiently slowly.  
Both the group speed of the wave 
and the radial component of the particle speed decrease as $1/R^2$ at
large $R$. Thus, even if the wave is launched with highly
relativistic particle streaming in its rest frame, this streaming
speed tends to zero at large $R$. This effect persists when damping 
is included in the computation, so that parametric instabilities will
set in at some stage and destroy the wave. This
point is the location of the termination shock.

\section{Conclusions}

The structure of the pulsar wind termination shock is determined by
the physical conditions not only in the magnetosphere, but also in the
external medium. Two regimes emerge from the model: the one with high
external pressure, in which the EM wave cannot be launched at all and
the shock forms rather due to interactions of the external medium with
the inner MHD wind; the second one is that with a lower external
pressure, in which case the EM wave exists as a stationary shock
precursor, which, after deceleration, becomes unstable and
leads to the formation of a shock front. The damping of the strong 
precursor wave by photon emission or inverse Compton scattering of external
photons can potentially provide an observable test of this scenario.

\acknowledgments 
IM would like to thank the IMPRS for Astronomy \& Cosmic Physics at
the University of Heidelberg for the financial support.
\bibliography{editor}
\end{document}